\documentclass[aps,amsfonts,amsmath,prd,preprint,tightenlines,nofootinbib,longbibliography]{revtex4-1}
\pdfoutput=1
\newcommand{\beq}{\begin{equation}}
\newcommand{\eeq}{\end{equation}}

\def\GW{{\text{GW}}}
\usepackage{graphicx,subfigure}
\usepackage[utf8]{inputenc}     

\def\nsf{\mathsf{n}}
\def\bea{\begin{eqnarray}}
\def\eea{\end{eqnarray}}

\usepackage{amsmath}

\begin{document}

\title{Comparison of cosmic string and superstring models to NANOGrav 12.5-year results}

\author{Jose J. Blanco-Pillado}
\affiliation{Department of Physics, University of the Basque Country, UPV/EHU, Bilbao, Spain}
\affiliation{IKERBASQUE, Basque Foundation for Science, 48011, Bilbao, Spain}
\author{Ken D. Olum}
\affiliation{Institute of Cosmology, Department of Physics and Astronomy,\\
Tufts University, Medford, MA 02155, USA }
\author{Jeremy M. Wachter}
\affiliation{Skidmore College Physics Department, 815 North Broadway, Saratoga Springs, \\
New York 12866, USA}

\def\changenote#1{\footnote{\bf #1}}

\begin{abstract} 
We compare the spectrum of the stochastic gravitational wave
background produced in several models of cosmic strings with the
common-spectrum process recently reported by NANOGrav.  We discuss
theoretical uncertainties in computing such a background, and show
that despite such uncertainties, cosmic strings remain a good
explanation for the potential signal, but the consequences for cosmic
string parameters depend on the model.  Superstrings could also
explain the signal, but only in a restricted parameter space where
their network behavior is effectively identical to that of ordinary
cosmic strings.
\end{abstract}
 
\maketitle
\thispagestyle{empty}
\section{Introduction}

The NANOGrav collaboration has recently reported some evidence of a
stochastic signal in their $12.5$-year data set on pulsar
timing~\cite{Arzoumanian:2020vkk}. Their observation of $45$ pulsars
indicates the presence of a common-spectrum ``red noise'' process. It
is unclear whether one can consider these results as a first hint of a
gravitational wave background in this frequency band. In particular,
the data so far show only weak evidence of a quadrupole
(Hellings-Downs \cite{Hellings:1983fr-special}) spatial correlation, so the
NANOGrav collaboration has not claimed a detection of a gravitational
wave signal yet. Further analysis is required in order to confirm this
as a first observation of an stochastic gravitational wave background
(SGWB). It is, however, tantalizing to consider this data seriously
and ask ourselves about its possible implications for astrophysics and
cosmology.

Our current understanding of galaxy evolution and merging history
leads us to the idea that there should be a large number of
supermassive black hole binaries (SMBHB) throughout the universe. This
incoherent sum of all such SMBHB will in turn produce an SGWB. The
predicted spectrum of this type of source in the nanohertz frequency
band has been estimated to be close to the current limits of the
Pulsar Timing Array (PTA)
observatories~\cite{Shannon:2015ect}. Moreover, the frequency
dependence of the spectrum is well known: the energy density in
such waves is given by a power law of the form
$\Omega^\GW_{\text{SMBHB}} \sim f^{2/3}$. All this makes them the most
likely candidate to explain a potential signal at these frequencies.

There are, however, other potential sources of gravitational waves at
these frequencies which are associated with cosmological processes in
the primordial universe. One of the most natural and promising sources
is the stochastic background of gravitational waves created by a
network of cosmic strings.  Cosmic strings are effectively
one-dimensional topological defects that may have been produced by a
phase transition in the early
universe~\cite{Kibble:1976sj,Vilenkin:2000jqa}. We will be interested
here in the simple case of Abelian-Higgs strings, or superstrings,
with no couplings to any massless particle other than the graviton.
Such a string network is described by a single quantity that
parameterizes the characteristic energy scale of the universe at the
time of string formation. This energy scale specifies the energy per
unit length of the string as well as its tension, $\mu$.  Since we are
interested in gravitational effects, we will be most interested in the
combination $G\mu$, where $G$ is Newton's constant.  We will work in
units where $c = 1$, so that $G\mu$ is dimensionless.

The equality between the energy per unit length of the strings and
their tension implies that the dynamics of these strings are
relativistic. Putting all of these facts together, one can immediately
see why cosmic strings are good candidates for gravitational waves:
they are cosmologically large relics that store very high energy
densities associated with the early universe, and they move
relativistically under their own tension. This explains why an
accurate computation of the SGWB from strings has been pursued for a
long time in the cosmic string
community~\cite{Vilenkin:1981bx,Hogan:1984is,Vachaspati:1984gt,Accetta:1988bg,Bennett:1990ry,Caldwell:1991jj,Damour:2004kw,Siemens:2006yp,DePies:2007bm,Olmez:2010bi,Sanidas:2012ee,Sanidas:2012tf,Binetruy:2012ze,Kuroyanagi:2012wm,Blanco-Pillado:2013qja,Sousa:2016ggw,Blanco-Pillado:2017oxo,Blanco-Pillado:2017rnf,Cui:2017ufi,Chernoff:2017fll,Ringeval:2017eww}.

Because string models are described by a single parameter related to the universe's energy at their time of formation, an observation of the SGWB from strings would indicate the existence of new physics at the string scale. However, the apparent simplicity of the single-parameter model is deceiving when it comes to detecting strings. The dynamics of the cosmic string network are complicated, making it difficult to obtain detailed descriptions of the necessary ingredients to compute the SGWB. One has to resort to large scale simulations to be able to establish basic facts needed in this calculation, like the number density of cosmic string loops throughout the history of the universe, or the typical power spectrum of such loops.\footnote{The string network contains both loops and long, horizon-spanning strings, but the contribution of long strings to the SGWB is subdominant for all $\mu$. We consider only the SGWB due to loops.} These are questions that one would have to answer in any model that produces an stochastic background: how many emitters are there, and how do they emit? Knowing this, we can estimate the combined effect of all sources. 

Comparisons of some cosmic string models' predictions with the
NANOGrav data have recently been made
in~\cite{Ellis:2020ena,Blasi:2020mfx,Buchmuller:2020lbh,Bian:2020bps}.
We will focus here on how theoretical uncertainties in the typical power
spectrum of a cosmic string loop impact the amplitude and slope of the
SGWB signal in the NANOGrav window, and therefore how the most-likely
$\mu$ (and associated confidence intervals) changes due to this
uncertainty. We do not suggest that a confirmed cosmic string
detection would resolve this theoretical uncertainty, as that requires
a better understanding of cosmic string networks and evolution.

\section{The SGWB from Cosmic Strings}\label{sec:cs-sgwb}

The basic idea behind the string SGWB computation is simple: for any given observational frequency, collect the contributions from all the different strings throughout the history of the universe that emit waves with the appropriate frequency such that they are observed at the observational frequency today. It is customary to present this information by calculating the critical density fraction of energy in gravitational waves per logarithmic frequency today,
\beq\label{eqn:cs-Omega}
\Omega_\GW (\ln f) = \frac{8 \pi G}{3 H_0^2} f \rho_\GW (t_0,f)\,,
\eeq
where $H_0$ is the Hubble parameter today, and $\rho_\GW$ denotes the
energy density in gravitational waves per unit frequency.

The calculation of the energy density has been described in detail
in~\cite{Blanco-Pillado:2017oxo}, and a summary can be found in
Appendix~\ref{app:sgwb}.  Each loop radiates in discrete multiples $n$
of its fundamental oscillation frequency $2/l$, where $l$ is the
invariant loop length, given by the loop energy divided by $\mu$.  We
write the power from loop $i$ in harmonic $n$ as $P_n^{(i)} G \mu^2 $,
so $P_n^{(i)}$ is dimensionless.  We write the total radiation power
$\Gamma^{(i)} G \mu^2$, where $\Gamma^{(i)} = \sum_{n=1}^{\infty}
P_n^{(i)}$.  For our purposes here, we will neglect differences in
$\Gamma^{(i)}$ between loops and just write $\Gamma$.

The three main ingredients we need to compute the string SGWB are:
\begin{itemize}
\item A cosmological model.
\item The number density of non-self-intersecting loops as a function of length at any moment in time.
\item The average power spectrum of gravitational waves from non-self-intersecting loops in the network, $P_n$.
\end{itemize}
We consider a standard cosmological history, and take the loop number
density described in~\cite{Blanco-Pillado:2017oxo} based on the
simulations reported in~\cite{BlancoPillado:2011dq}.  This leaves the
average power spectrum of non-self-intersecting loops, $P_n$. This is
probably the quantity in the calculation with the highest uncertainty
at this moment, since it depends not only on the gravitational
radiation spectrum of non-self-intersecting loops at formation, but
also on their evolution. This is a challenging problem, since one
needs to follow the change in shape of a representative set of
non-self-intersecting loops throughout their lifetimes; in other words,
one needs to account for gravitational backreaction. Lacking this
information, one can either take an ansatz for backreaction, or model
the power spectrum in some theoretically-motivated way which should
hold true, in general, even after accounting for backreaction.

An early attempt to take gravitational backreaction into account was
done in~\cite{Blanco-Pillado:2015ana}. There, the authors implemented
a toy model for backreaction on a large set of non-self-intersecting
loops obtained from the simulations described
in~\cite{BlancoPillado:2011dq}. The idea behind this toy model was to
simulate backreaction by smoothing structures on the loops at
different time scales. The results of this procedure indicated that the
distribution of values of the total power was peaked around
$\Gamma\sim 50$, which we will take as the $\mathrm{\Gamma}$ for all
SGWB we study.  We use $P^{BOS}_n$ (after the authors' initials) to
indicate the average power spectrum computed by this work, and will
use it as one of the models we study in the following section. It is
quite smooth, and has a long tail describing the emission of a
substantial amount of power at the high-frequency modes of the
string. This can be traced to the presence of cusps in the final
stages of the evolution of these smoothed loops.  The SGWB spectra
arising from this model were discussed in~\cite{Blanco-Pillado:2017oxo}.

Cusps are moments of the loop's oscillation when a point on the loop
formally reaches the speed of light~\cite{Turok:1984cn}. Cusp
formation leads to the loop emitting a significant amount of
radiation, which is beamed in the direction of motion of the
cusp~\cite{Vachaspati:1984gt}.\footnote{Cusp bursts can be sources of
  transient events in gravitational wave detectors. See the discussion
  in~\cite{Damour:2000wa,Damour:2001bk,Damour:2004kw}.} Accumulating
radiation from many such events forms a stochastic background whose
power spectrum has a long tail, of the form $P^{\text{cusp}}_n \propto
n^{-4/3}$~\cite{Vachaspati:1984gt}. Because cusps are thought to be
generic features of loops, a common model of the power spectrum is one
where low modes, which describe the shape of the loop, are less
important than high modes. If we focus on these high-mode
contributions to gravitational waves, then we can use a model where
the spectrum is simply given by $P^{\text{cusp}}_n$.  We will choose a
constant of proportionality so that $\sum_{n=1}^{\infty}
P^{\text{cusp}}_n = \Gamma$.  This is the second model of $P_n$ we
consider when discussing a possible string SGWB.

Another characteristic feature on realistic loops are kinks: points
along the string where there is a discontinuity in its tangent
vector. These occur every time two segments of string intersect one
another and exchange partners. Kinks move at the speed of light along
the string, emitting a fan of radiation whose spectrum at high mode
number emission goes as $P^{\text{kink}}_n \propto
n^{-5/3}$~\cite{Damour:2001bk}. As with cusps, we can consider a model
with only kink radiation. This is our third model.

A fourth and final model takes the reverse approach: instead of
focusing on the high-harmonic tail, we consider a spectrum consisting
only of the fundamental mode,
\beq
 P_n^{\text{mono}} = \left\{\begin{array}{ll} \mathrm{\Gamma} & \text{if}\quad n=1\\ 0 & \text{otherwise}\end{array}\right.\,.
\eeq
Like the pure-cusp and pure-kink spectra, this is not a realistic
assumption, but it serves as a limiting case for strings which radiate
primarily in low harmonics.

The real average spectrum should be calculated from a realistic
distribution of non-self-intersecting loops obtained from a scaling
simulation \emph{and} evolved under their own gravity. This can be
done using linearized gravity, since the force that affects each
loop's shape depends on $G\mu$, which in our case is always very
small. This idea was first developed in~\cite{Quashnock:1990wv}, and
has recently been advanced both
analytically~\cite{Wachter:2016rwc,Wachter:2016hgi,Blanco-Pillado:2018ael,Chernoff:2018evo}
and numerically~\cite{Blanco-Pillado:2019nto}. The results from these
papers indicate that cusps and kinks are smoothed over time. Some of
the effects of backreaction are captured by the smoothing procedure
of~\cite{Blanco-Pillado:2015ana}, but there are cases where this
approach is not so accurate. The specific results of the long-term
effect of backreaction on loops produced by a scaling string network
are therefore still unclear.  Thus we will show the gravitational wave
amplitudes and spectral slopes to be expected for all four models and
compare them with the NANOGrav observations.

\section{Comparison with NANOGrav 12.5-year data}

The NANOGrav collaboration presents their data using the characteristic strain of the form
\beq
h_c(f) = A~\left(\frac{f}{f_{\text{yr}}}\right)^{\alpha} = A~\left(\frac{f}{f_{\text{yr}}}\right)^{(3 - \gamma)/2}\,,
\eeq
where $f_{\text{yr}} = 1/\text{year}$, $A$ is the strain amplitude, and $\gamma$ is the spectral index. The energy density in gravitational waves can be obtained from this characteristic strain using the relation
\beq
\Omega_\GW (f) = \frac{2 \pi^2}{3 H_0^2} f^2 h_c^2(f)\,.
\eeq
NANOGrav also reports likelihoods in the parameter space of $(\gamma,A)$, which we will use to construct confidence regions to use for our analysis of the effect of different $P_n$. 

For a given $G\mu$ and $P_n$, we can compute the energy density in
gravitational waves with Eq.~(\ref{eqn:cs-Omega}). From this, we
approximate the spectral index and amplitude using the two lowest
frequencies seen in NANOGrav, $f_1=1/(12.5\,\text{yr})$ and
$f_2=2f_1$. This process provides a good fit to compare to the
5-frequency contours because the two lowest frequencies in NANOGrav
are much better determined in comparison to the third through fifth lowest frequencies. Our method is to calculate
\begin{subequations}\begin{align}
    \gamma &= 5-\frac{\ln(\Omega_\GW(f_2)/\Omega_\GW(f_1))}{\ln(2)}\,,\label{eqn:gamma}\\
    A &= \sqrt{\frac{3H_0^2 \Omega(f_1) f_\text{yr}^{3-\gamma}}{2 \pi^2 f_1^{5-\gamma}}}\label{eqn:A}
\end{align}\end{subequations}
for each $\Omega_\GW$.

\begin{figure}
\centering
\includegraphics[scale=1.0]{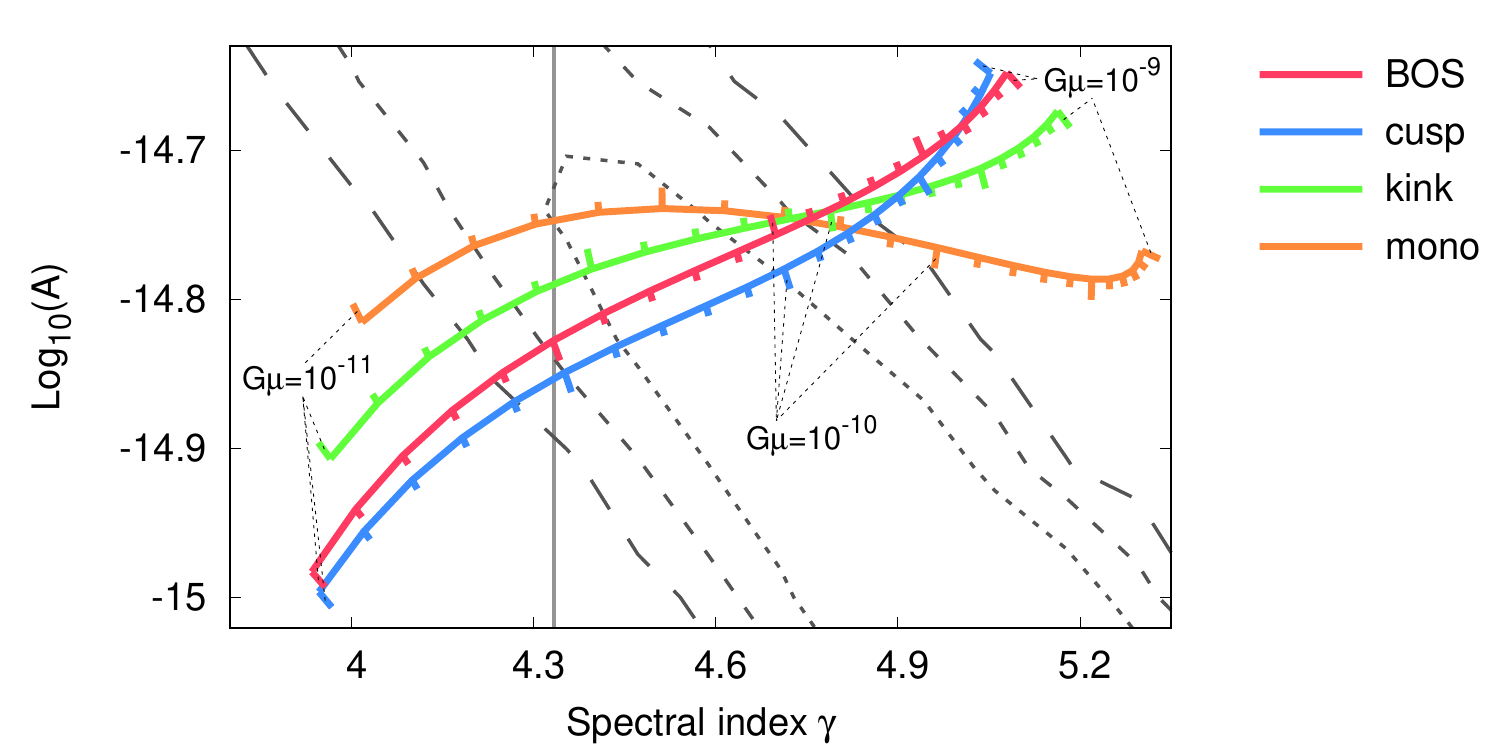}
\caption{The amplitude vs. spectral index for various cosmic string
  tensions, $G\mu$, for four models of the average power spectrum,
  $P_n$. The tic marks show steps of $0.1$ in $\log G\mu$, from $-9$
  to $-11$, with large tics every $0.5$.  The short, medium, and long
  dashes show the $1\sigma$, $1.5\sigma$, and $2\sigma$ contours (i.e.,
  they enclose 68\%, 86\%, and 95\% of the likelihood, respectively)
  made from the NANOGrav $12.5$-yr 5-frequency chain
  data~\cite{NANOGrav-data}. We follow the process outlined in the PTA
  GWB Analysis tutorial~\cite{NANOGrav-code} and then extract the
  contour control points directly from the resulting graphical
  object~\cite{corner}. We use 50 bins in each direction to produce
  higher-resolution contours than are seen in Fig.~1 of 
  \cite{Arzoumanian:2020vkk}. The vertical gray line shows the
  spectral index to be expected from SMBHB.}
\label{fig:allPn}
\end{figure}

Figure~\ref{fig:allPn} shows the curves one obtains in the $(\gamma,
A)$ plane for $G\mu\in[10^{-9}, 10^{-11}]$ for our four models of
$P_n$. All models have been normalized so that the total power is
given by $\mathrm{\Gamma}= 50$.
We report the approximate ranges of $\log G\mu$ which predict values within the $1\sigma$, $1.5\sigma$, and $2\sigma$ confidence range in Table~\ref{tbl:best-Gmu}.

\begin{table}
\centering
\begin{tabular}{l|l|l|l}
\textbf{$P_n$ model} & \textbf{$1\sigma$ range} & \textbf{$1.5\sigma$ range}  & \textbf{$2\sigma$ range}\\
\hline
BOS  & $(-10.08,-10.40)$ &  $(-9.92,-10.52)$ &  $(-9.77,-10.62)$\\
cusp & $(-10.02,-10.39)$ &  $(-9.85,-10.50)$ &  $(-9.72,-10.60)$\\
kink & $(-10.24,-10.52)$ & $(-10.08,-10.64)$ &  $(-9.93,-10.74)$\\
mono & $(-10.45,-10.67)$ & $(-10.27,-10.80)$ & $(-10.08,-10.90)$\\
\end{tabular}
\caption{The approximate values of $\log G\mu$ falling within the
  $1\sigma$, $1.5\sigma$, and $2\sigma$ confidence intervals of
  NANOGrav for the four $P_n$ models.}\label{tbl:best-Gmu}
\end{table}

The important general result of Fig.~\ref{fig:allPn} is that the $(\gamma, A)$ parameters predicted by the different spectra are quite similar over the range we investigate. This means that the theoretical uncertainty in the average gravitational wave power spectrum from loops will not greatly affect the conclusions obtained from identifying the NANOGrav result with the SGWB from cosmic strings. In other words, assuming the actual spectrum of realistic loops is somewhere close to the models we study here, we can infer that the constraints on $G\mu$ are quite similar to the ones obtained from this figure. Of course, future data and analysis will likely reduce uncertainties, shrinking the range of the significance contours, and allowing us to pin down the most likely value of $G\mu$. Our ability to do that will depend on reducing our uncertainty in the loop power spectrum. This is the job of simulation, and a detection consistent with any of the above curves should not be considered evidence for that $P_n$ being the true power spectrum of loops in nature.

\section{Relationship to previous work}

\subsection{Upper bounds}

The authors of~\cite{Blanco-Pillado:2017rnf}, including two of
us, derived bounds on the possible values of $G\mu$ from
non-observation of a SGWB.  We concentrated on the BOS model.  Using
results from the Parkes PTA \cite{Shannon:2015ect,Lasky:2015lej}, we
gave a limit of $G\mu < 1.5\times 10^{-11}$, and using the the
NANOGrav 9-year results~\cite{Arzoumanian:2015liz}, we gave $G\mu <
4.0\times 10^{-11}$.  However, referring to Fig.~\ref{fig:allPn}, we
see that the best fit $G\mu$ is about $5.6\times 10^{-11}$, 40\% 
larger than the limit based on NANOGrav and about 4 times the limit based
on Parkes.

There are two reasons for this discrepancy.  First, all pulsar timing
arrays include models of individual pulsar noise.  If not treated
correctly, this modeling can absorb the effects of the SGWB, leading
to incorrect upper bounds.  This is discussed in detail in
\cite{Hazboun:2020kzd}. In~\cite{Arzoumanian:2020vkk}, the authors
compare the NANOGrav red noise process detection with their
previously given upper limits.

Second, pulsar timing is dependent on the solar system ephemeris,
which tells us how to remove the earth's motion through the solar system
from the observed data.  We do not know this ephemeris to the accuracy
necessary, and thus ephemeris uncertainty is an additional source of
error in gravitational wave measurements.  In particular, if one
allows the observations to influence the choice of ephemeris, one may
thereby absorb some gravitational wave power and infer incorrect
limits.  See~\cite{Arzoumanian:2018saf} for more detailed
discussion.

\subsection{Other cosmic string SGWB results}

Other recent papers~\cite{Ellis:2020ena,Blasi:2020mfx,Buchmuller:2020lbh,Bian:2020bps} have interpreted the NANOGrav 12.5-year data as a cosmic string signal. We discuss the similarities and differences between their approaches and ours here, and comment generally on agreements between those approaches.

The majority of the sources mentioned~\cite{Ellis:2020ena,Blasi:2020mfx,Bian:2020bps} employ the velocity-dependent one-scale (VOS) model for generating the cosmic string SGWB. This model has an additional parameter: the loop size at formation as a fraction of horizon size, $\alpha$, which~\cite{Ellis:2020ena} sets to $0.1$ and which~\cite{Blasi:2020mfx,Bian:2020bps} allow to vary over some range.\footnote{When not exploring the effect of varying $\alpha$, taking $\alpha=0.1$ for the VOS model is a typical one, based on simulations of string networks.} Reference~\cite{Buchmuller:2020lbh} follows the same approach as this paper. All of the aforementioned use a cusp power spectrum in creating their SGWB, and so we can only make meaningful comparisons between their results and our cusp results.

The VOS model and the one we use here are in near-exact agreement when
VOS takes $\alpha=0.1$ and one corrects for the overall energy loss
into kinetic energy of the loops~\cite{Auclair:2019wcv}. We would
therefore expect close agreement between our results and those of
~\cite{Ellis:2020ena}, and between our results and those of
~\cite{Blasi:2020mfx,Bian:2020bps} for
$\alpha=0.1$.\footnote{Note that~\cite{Blasi:2020mfx,Bian:2020bps}
conclude that values of $\alpha<0.1$ produce better fits to the NANOGrav data.}
Reference~\cite{Buchmuller:2020lbh} considers metastable cosmic
strings, characterized by a parameter $\kappa$; we would expect their
results to match ours in the limit $\kappa\rightarrow\infty$, i.e.,
when the decay rate of cosmic strings due to monopole--antimonopole
pair production goes to zero and the strings decay only via GWs.

There is one additional concern in comparing different results in the
$\gamma$-$\log A$ plane.  Suppose two different approaches generate
identical SGWB, so they predict the same $\Omega_\GW$ at some common
reference frequency $f_\text{ref}$, but they use different approaches
to determine $\gamma$.  When they extrapolate the amplitude from
$f_\text{ref}$ to $f_\text{yr}$ to report $A$ (see Eq.~(\ref{eqn:A})),
the resulting $A$ will be different.  The difference in the reported
logarithmic amplitude is
\beq
\Delta(\log(A)) = \log(A_1/A_2) = \frac12(\gamma_2-\gamma_1)\log\left(\frac{f_\text{ref}}{f_\text{yr}}\right)\,.
\eeq

Taking this effect into account,~\cite{Ellis:2020ena} draws very
similar conclusions to ours as to the bounds on $G\mu$, as does
~\cite{Blasi:2020mfx} for the $\alpha=0.1$ case.
Reference~\cite{Bian:2020bps} does not display their results for
$\alpha=0.1$, and so we cannot make a direct
comparison. Reference~\cite{Buchmuller:2020lbh} does not display a
comparison to their results with a stable string SGWB, but their
bounds on $G\mu$ as $\kappa$ increases seem to be converging towards
results consistent with ours (e.g., the point with $G\mu=10^{-10}$ and
largest $\kappa$ is on the edge of the $1\sigma$ contour).

\section{Cosmic Superstrings}

Until now, we have been discussing the gravitational spectrum produced
by a network of cosmic strings that exchange partners whenever they
intersect. This is the expected interaction of strings that appear as
topological defects in field theory (e.g., in the Abelian-Higgs
model~\cite{Shellard:1987bv,Matzner:interaction}). There are, however,
other scenarios where a network of cosmologically interesting
string-like objects is produced. In particular, many cosmological
models of superstring theory suggest the production of fundamental
strings, which are then stretched to cosmological size by an expanding
universe~\cite{Sarangi:2002yt,Dvali:2003zj,Copeland:2003bj,Polchinski:2004ia}. Once
stretched, these fundamental strings have similar dynamics to their
classical counterparts, except for the crucial aspect that their intercommution is
different. This is due to the fact that their interactions are 
quantum mechanical in origin, and also because the strings in these models
 may move in a space with additional dimensions.  Both these
effects may significantly reduce their chance to intercommute. That
is, the strings sometimes pass through one another, rather than
splitting and rejoining to form sharply-angled kinks. This issue has
been studied in~\cite{Jackson:2004zg}, where the conclusion was that
the probability $p$ of reconnection could be as low as $10^{-3}$.

A decrease in the intercommutation probability should have an effect
on the macroscopic properties of the network. There has been some
debate in the literature about how this lower probability would modify
the overall density of the
strings~\cite{Dvali:2003zj,Sakellariadou:2004wq,Avgoustidis:2005nv}. This
is important to the calculation of the SGWB, since the density of
loops has a direct impact on the size of $\Omega_\GW$. Large scale
simulations would be necessary to establish the precise modifications
that this reduced probability will bring to the final scaling
distribution of loops presented earlier, but they have not yet been
done. Here, we will assume that the effect of reducing $p$ is to
increase the loop number density by factor $1/p$, without changing the
properties of the loops, so that
\beq
\Omega_\GW \propto \frac1p\,.
\eeq

\begin{figure}
\centering
\includegraphics[width=15.0cm]{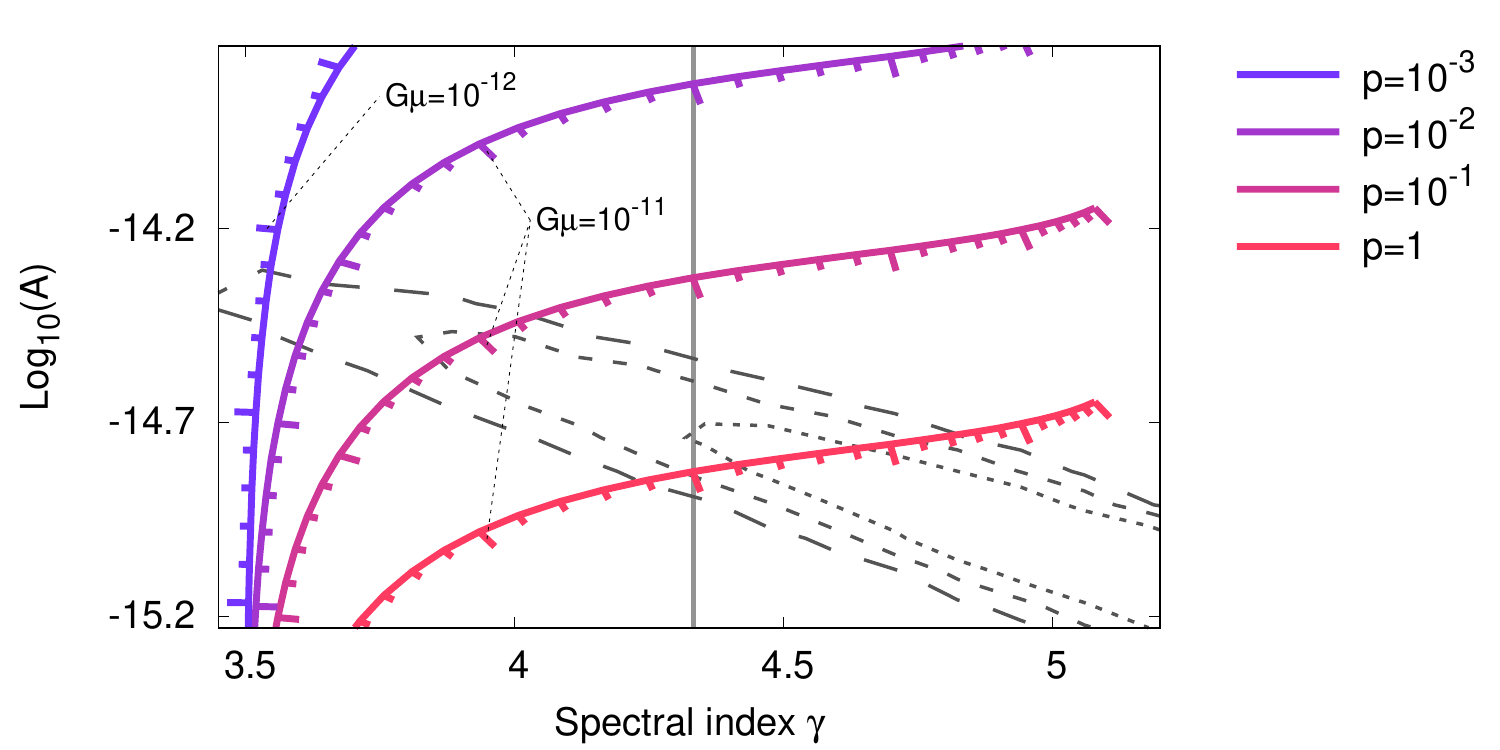}
\caption{The amplitude vs. spectral index for various superstring tensions, $G\mu$, for the BOS model of the average power spectrum. The intercommutation probabilities go from $1$ to $10^{-3}$ in powers of ten, with lower $p$ moving the curve out of NANOGrav's significance region (c.f. Fig.~\ref{fig:allPn}), represented by the dashed grey lines. The other models of $P_n$ return similar results. The vertical gray line shows the spectral index to be expected from SMBHB.}
\label{fig:superstrings}
\end{figure}

Lowering the intercommutation probability increases the amplitude of
gravitational waves without changing the slope, and so we may estimate
the range of $p$ which is compatible with the current NANOGrav
data. The upward displacement of the curves in the $(\gamma,A)$ plane
quickly moves them away from the $1\sigma$ region, as shown in
Fig.~\ref{fig:superstrings}, in agreement with the result of
~\cite{Ellis:2020ena}.

However, the current likelihood data will never completely exclude a
superstring network at the $2\sigma$ level.  The $1/p$ enhancement means
that for small $p$ we are interested in a smaller $G\mu$. This puts us
in the low-$f$ region in the cosmic string background spectrum
\cite{Blanco-Pillado:2017oxo}, where $\Omega_\GW$ rises with frequency
as $f^{3/2}$, giving $\gamma=7/2$.  For any small $p$, there will be
some $G\mu$ giving the $A$ that lies in the $2\sigma$ region at the left
of Fig.~\ref{fig:superstrings}.  While we only show superstrings using
the BOS model of $P_n$, the $3/2$ rising slope does not depend on $P_n$,
and so this effect is generally true.

Despite this, the NANOGrav data as currently given is most consistent
with $p\approx 1$. As a consequence, superstrings are likely to explain
the potential signal only if their network properties are very similar to
those of cosmic strings.

A counterargument to this claim is the idea that strings are wiggly, and so each string crossing has multiple potential intersection events, increasing the chance that strings intercommute and thus depressing the $1/p$ enhancement to the energy density. A specific example of such an argument can be made using the results of~\cite{Avgoustidis:2005nv}, which found the energy density to have very little enhancement down to $p\approx 0.1$, after which it follows $\Omega_\GW\propto p^{-0.6}$. This would relax the bounds on $p$ somewhat, allowing superstrings down to $p\sim 10^{-2}$ to fall at the edge of the $1.5\sigma$ region, roughly where $p=10^{-1}$ lies in Fig.~\ref{fig:superstrings}.

Our conclusions about superstring viability change slightly if
improved statistics moves the confidence interval contours towards the
left, towards the predicted SMBHB signal's vertical line at
$\gamma=13/3$. There, an enhancement to the amplitude due to $p\sim
0.1$ would cause the string curves presented to overlap with the SMBHB
signal around $G\mu\lesssim 10^{-10.5}$. It may therefore be necessary
to distinguish a superstring SGWB from a supermassive black hole
binary SGWB. Because we expect the number of cosmic strings or
superstrings that contribute to the SGWB to be large in the frequency
band seen by NANOGrav, this could be accomplished by studying
anisotropies in the reported signal, which we would not expect if
strings are the source.

\section{Conclusions}

Regardless of the model chosen to represent the average power spectrum
of a cosmic string loop, the potential signal reported by NANOGrav
could be a cosmic string stochastic gravitational-wave
background. Thus, as long as these models are close to the true
average power spectrum, a confirmation of a cosmic-string signal would
predict the existence of a network of strings with a tension in the
range of $G\mu\approx [10^{-10.0},\,10^{-10.7}]$. Such $G\mu$ values are
low enough that we would not expect such strings to be visible in the 
cosmic microwave background~\cite{Ade:2013xla} or to produce 
gravitational wave bursts that can be seen in 
interferometers~\cite{Abbott:2019prv}\footnote{See also the results presented 
for model A in \cite{Abbott:2021ksc}.} or pulsar timing 
arrays~\cite{Yonemaru:2020bmr}.

Superstrings are less favorable as an explanation for the signal. They
would either have to have very similar network properties to cosmic
strings, due to $p\approx 1$, or would have to be rescued by changes
to the confidence interval contours.

\begin{figure}
\centering
\includegraphics[width=15.0cm]{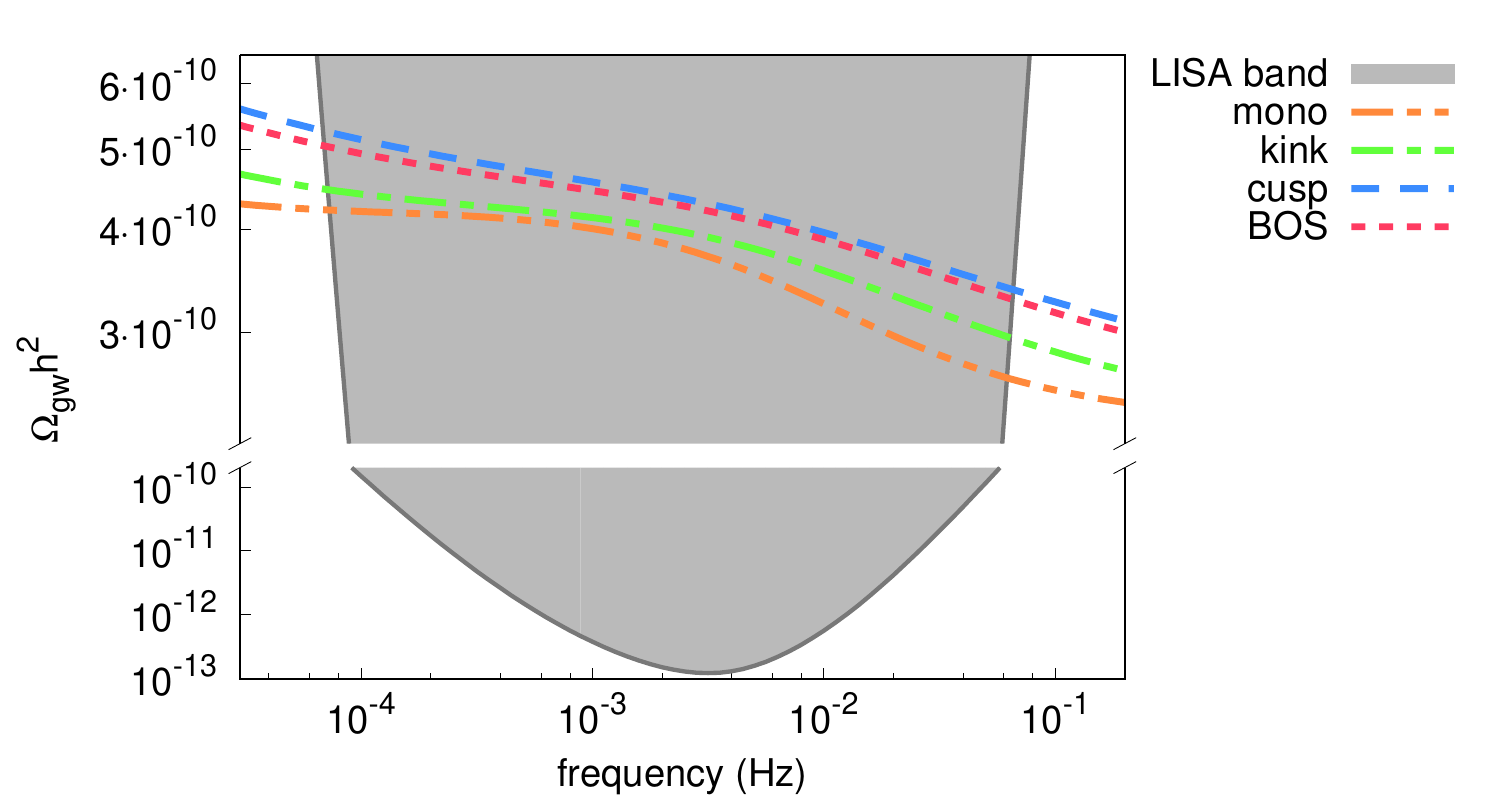}
\caption{The energy density vs. frequency of the SGWB for the four
  $P_n$ models we consider, as seen in the LISA band. All curves are
  for $G\mu=10^{-10}$, but the range of tensions which fit the
  NANOGrav data produce similar results. The decline in the lines, and
  its variation, are direct consequences of changing degrees of
  freedom in the universe's past, and so LISA could measure deviations
  from a standard cosmological model for such
  curves~\cite{Caprini:2019pxz}.}\label{fig:allPnLISA}
\end{figure}

If the signal is indeed from cosmic strings, then we can expect to see
other parts of the SGWB in future gravitational wave telescopes. The
values of $G\mu$ we consider are too low for LIGO/VIRGO to observe the
SGWB~\cite{Abbott:2021ksc}\footnote{Here we only consider model A in~\cite{Abbott:2021ksc}.
See~\cite{Blanco-Pillado:2019vcs} for a critical discussion of the viability of other models presented in this reference.},
but LISA, the Einstein Telescope, or the BBO are sensitive in
the correct frequency and amplitude range. In LISA, for example, we
could measure the section of the SGWB which contains information about
cosmological history, particularly the effect of changing degrees of
freedom
\cite{Blanco-Pillado:2017oxo,Cui:2018rwi,Auclair:2019wcv,Caprini:2019pxz},
as shown in Fig.~\ref{fig:allPnLISA}. Such a measurement could be used
to quantify deviations from the standard model and thus probe new
physics.

\section*{Acknowledgments}

We would like to thank Xavier Siemens for helpful conversations.  This
work is supported in part by the Spanish Ministry MCIU/AEI/FEDER grant
(PGC2018-094626-B-C21), the Basque Government grant (IT-979-16) and
the Basque Foundation for Science (IKERBASQUE), and in part by the
National Science Foundation under grant number 1820902.

\appendix

\section{Computing the gravitational wave energy density}\label{app:sgwb}

Fundamental to Eq.~(\ref{eqn:cs-Omega}) is $\rho_\GW$, the energy density in gravitational waves per unit frequency. It can be written
\beq\label{eqn:cs-rho}
\rho_\GW (t,f) = G \mu^2 \sum_{n=1}^{\infty} C_n P_n\,,
\eeq
where $P_n$ describes the average gravitational wave spectrum of the cosmic string loops in the network and
\beq\label{eqn:Cn}
C_n = \int_0^{t_0} \frac{dt}{(1+z)^5} \frac{2n}{f^2}~\nsf (l,t) = \frac{2n}{f^2} \int_0^{\infty} {\frac{dz}{H(z) (1+z)^6} \nsf\left( \frac{2n}{(1+z)f},t(z)\right) }\,,
\eeq
where $\nsf(l,t)$ is the loop number density, $H(z)$ is the Hubble
parameter and $t(z)$ the age of the universe at redshift $z$.  We
consider a standard cosmological history, so $H(z)$ and $t(z)$ are
given by the usual expressions in terms of the components of
the universe, $\Omega_r$, $\Omega_m$, and $\Omega_{\Lambda}$, as well as
the number of degrees of freedom at each moment in time
(see~\cite{Blanco-Pillado:2017oxo} for a detailed explanation of these
functions).

Finding the form of $\nsf (l,t)$, is equivalent to finding the
distribution of non-self-intersecting loops at all times in the
history of the universe. This sounds like a challenging problem since
it will be impossible to simulate the evolution of the network for
such a wide range of time scales. Luckily for us the evolution of a
cosmic string network has a {\it scaling solution}, where the energy
density of the string remains a small fraction of the background
energy density of the universe. This is an important property of the
model since it makes cosmological string networks compatible with
observations. There is a more important aspect of this scaling
solution for our calculation: in a scaling solution, the form of the
loop distribution satisfies
\beq
\nsf (l,t) = t^{-4} \nsf(x)\,,
\eeq
where $x = l/t$ is the ratio of the loop size to the age of the universe at some particular time, and $\nsf (x)$ is the number of loops per unit $x$ in a volume $t^3$. The scaling solution simplifies the problem, reducing it to finding $\nsf(x)$. Finding this scaling solution from numerical simulations presents a big challenge, since one has to run for extremely long periods of time before reaching a true scaling solution for the loop distribution.\footnote{See, for example, \cite{Blanco-Pillado:2019tbi} for a discussion of the existence of transient solutions early in a simulation.} Here, we use the results of the Nambu-Goto simulations presented in~\cite{BlancoPillado:2011dq} and analyzed in~\cite{Blanco-Pillado:2013qja}, which allow us to write the distribution for loops as
\beq
\nsf_r(l,t) = \frac{0.18}{t^{3/2}\left(l + \mathrm{\Gamma} G \mu t \right)^{5/2}}
\eeq
for loops existing in the radiation era. Some of these loops will
survive until the matter era, when they will contribute to the number
of loops as
\beq
\nsf_{rm}(l,t) = \frac{0.18\left(2\sqrt{\mathrm{\Omega}_{r}} \right)^{3/2}}{(l +\mathrm{\Gamma} G \mu t)^{5/2}}\left(1+z\right)^3\,.
\eeq
Finally, loops produced in the matter era contribute as
\beq
\nsf_m(l,t) = \frac{0.27 - 0.45 (l/t)^{0.31}}{t^2 (l +\mathrm{\Gamma} G \mu t)^2}\,,
\eeq
for $l<0.18t$, but these make no significant contribution to the
gravitational wave spectrum today.

We note that these expressions depend on the parameter $\mathrm{\Gamma}$, which describes the average total power of gravitational radiation emitted by the population of non-self-intersecting loops. The emission of energy into gravitational waves reduces the length of the loop according to
\beq
l = l_0 - \mathrm{\Gamma} G \mu (t - t_0)\,,
\eeq
which is why the previous expressions depend on $\mathrm{\Gamma}$. 

The final ingredient, $P_n$, is discussed in the main text.

\bibliography{NANOGrav-and-strings,no-inspire}

\end{document}